\pdfoutput=1
\documentclass[aps, reprint, superscriptaddress, nofootinbib, showpacs,preprintnumbers,floatfix]{revtex4-1}
\usepackage{amsmath, latexsym, amssymb, hyperref, graphicx, color, multirow}
\usepackage[table]{xcolor}
\usepackage{tabularx}
\usepackage[T1]{fontenc}
\usepackage[capitalize]{cleveref}
\usepackage{booktabs} 
\usepackage{fouriernc} 
\definecolor{nicered}{rgb}{.7,.1,.1}
\definecolor{nicegreen}{rgb}{.1,.5,.1}
\definecolor{darkblue}{rgb}{0,0,.5}
\hypersetup{colorlinks, citecolor=nicegreen,linkcolor=nicered, urlcolor=darkblue}

\begin{document}
\preprint{ACFI-T20-17}

\title{Left-right symmetry and electric dipole moments. A global analysis}
\author{Michael J. Ramsey-Musolf}
\email{mjrm@physics.umass.edu}
\affiliation{Tsung-Dao Lee Institute and School of Physics and Astronomy, Shanghai Jiao Tong University, 800 Dongchuan Road, Shanghai, 200240 China.}
\affiliation{Amherst Center for Fundamental Interactions, Department of Physics, University of Massachusetts, Amherst, MA 01003, USA.}
\affiliation{Kellogg Radiation Laboratory, California Institute of Technology, Pasadena, CA 91125 USA.}
\author{Juan Carlos Vasquez}
\email{jvasquezcarm@umass.edu}
\affiliation{Amherst Center for Fundamental Interactions, Department of Physics, University of Massachusetts, Amherst, MA 01003, USA.}


\begin{abstract}
\noindent
We perform a global fit using results of searches for electric dipole moments (EDM) of diamagnetic systems within the context of the minimal left-right symmetric model. In this way, we disentangle the new \lq\lq left-right'' electroweak and  $\bar{\theta}$ contributions that cannot be separated using a single EDM system. Although the fit is done for a specific model, the approach can be applied to any particle physics model. Finally, we revisit the constraint on  the $D$ coefficient in $\beta$-decay and find that current EDM bounds do not preclude observation of this T-violating effect in a possible next generation $\beta$-decay experiment. 
\end{abstract}

\maketitle

\section{Introduction}

The Standard Model (SM) of fundamental interactions contains two sources of CP violation (CPV): the $\bar{\theta}$ parameter and the phase $\delta$ in the Cabibbo-Kobayashi-Maskawa (CKM) mixing matrix. The $\bar{\theta}$ parameter is written as $\bar{\theta}=\theta_0+\text{arg det}(M_uM_d)$, where $\theta_0$ is the parameter governing the strength of the $G{\tilde G}$ interaction in the QCD Lagrangian and $M_u,M_d$ are the up and down quark mass matrices. Barring accidental cancelations, electric dipole moment (EDM) systems are in general sensitive to any source of flavor-diagonal CP violation. In particular $\bar{\theta}$ is the physical quantity entering in EDM expressions~\cite{PhysRevD.19.2227,Crewther:1979pi}. Consequently, one cannot disentangle the $\theta_0$ contribution from the $\text{arg det}(M_uM_d)$ contribution within the SM. Notice that, if the $\bar{\theta}$ parameter is zero at the tree level, radiative corrections within the SM turns out to be small and $\bar{\theta}\sim 10^{-19}$~\cite{Ellis:1978hq,Khriplovich:1985jr}, well below present experimental EDM bound: $\theta_{exp}\sim 10^{-10}$~\cite{PhysRevLett.119.119901,Abel:2020gbr} --see Ref.~\cite{Ginges:2003qt,Pospelov:2005pr,Engel:2013lsa,Chupp:2017rkp} for reviews. 

New physics beyond the Standard Model (BSM) may introduce new sources of CPV, which can then be the dominant source of CPV in hadronic, atomic, and molecular EDM systems. To illustrate, consider the neutron EDM, which has the general form\cite{Engel:2013lsa}
%
\begin{equation}\label{dn}
 d_n = \alpha_n\, \bar{\theta} +\beta_n^j\, \left(\frac{v^2}{\Lambda^2}\right)\sum_j\mathrm{Im}\left(C_j\right)+\cdots, 
\end{equation}
where $v=246$ GeV is the electroweak scale; $\Lambda$ is the BSM scale; $C_j$ are Wilson coefficients of dimension-six ($d=6$) operators built from SM fields~\footnote{The dimension-six sources include the quark and lepton EDMs, quark chromo-EDMs, CPV 3-gluon operator, and a set of four fermion operators. For a summary, see Ref.~\cite{Engel:2013lsa}.}; the coefficients $\alpha_n$ and $\beta_n^j$ encode the sensitivity of $d_n$ to $ \bar{\theta}$ and the BSM CPV sources, respectively; and the \lq\lq $+\cdots$\rq\rq denote contributions from higher-dimensional operators. Analogous expressions apply to other EDM systems, with correspondingly different sensitivities to ${\bar\theta}$ and the BSM CPV sources. Notice that if only one EDM system is probed, there is no way of disentangling the $\bar{\theta}$ and BSM contributions. In this case, no rigorous bound can be given either to $\bar{\theta}$ or $\mathrm{Im}(C_j)/\Lambda$ without making additional assumptions. However, results from a variety of EDM searches on systems with complementary sensitivities can yield discriminatory power.


%
In this work, we consider the implications of EDM searches in the context of the minimal left-right symmetric model (mLRSM)~\cite{Pati:1974yy,Mohapatra:1974gc,Senjanovic:1975rk,SENJANOVIC1979334,Mohapatra:1979ia,Mohapatra:1980yp}. In particular, we focus on diamagnetic systems, for which the leading mLRSM contributions are ${\bar\theta}$ and the $d=6$ \lq\lq left-right" operator defined below (see Refs.~\cite{Engel:2013lsa,Seng:2014pba} for a discussion of the various $d=6$ operators in the mLRSM). We perform a global analysis of diamagnetic EDM search results, following the approach taken in Refs.~\cite{Chupp:2014gka,Chupp:2017rkp} that considered the low-energy effective parameters relevant to the EDM systems. 

The implications of the EDM results depend decisively on whether parity ($\mathcal{P}$) or charge conjugation ($\mathcal{C}$) is adopted as the LR symmetry. For $\mathcal{P}$, one has $\theta_0= 0$ and $\bar{\theta}=\text{arg det}(M_uM_d)$ is calculable in terms of the mLRSM parameters. The latter is also true for other solutions of the ``strong CP problem", such as the soft breaking of parity~\cite{Mohapatra:1978fy,PhysRevLett.41.278,Mohapatra:1982ib,PhysRevD.41.1286,PhysRevLett.67.2765} or the soft breaking of CP symmetry~\cite{Georgi:1978xz,PhysRevLett.41.278}.  In each case, $\bar{\theta}$ is a calculable parameter, leaving only $\mathrm{Im}(C_j)/\Lambda^2$ is the {\em a priori} unknown quantity.  

This is not the case in general -- including the case of the SM -- since there is no way of disentangling the $\theta_0$ contribution from the radiative corrections giving rise to $\text{arg det}(M_uM_d)$. For $\mathcal{C}$ as the LR symmetry, $\theta_0\neq0$, and hence no bound on $\mathrm{Im}(C_j)/\Lambda^2$ can be obtained using a single EDM system. This is our main motivation for considering the set of diamagnetic systems. Doing so allows us to give bounds on both $\bar{\theta}$ and $\mathrm{Im}(C_j)/\Lambda^2$ without making any additional theoretical assumptions. 

In this context, we find -- unsurprisingly -- that EDM limits on $\bar{\theta}$ and $\mathrm{Im}(C_j)/\Lambda^2$ are significantly relaxed compared to a \lq\lq sole source" analysis in which only one of the two is assumed to be present. Perhaps, more interesting are the implications for the possible manifestations of CPV in other observables. In particular, we consider the T-odd correlation neutron $\beta$-decay, which has recently been constrained by the emiT collaboration~\cite{Mumm:2011nd,Mumm:2011zz}. We show that present EDM constraints would allow for observation of a non-zero effect in a future experiment performed with improved sensitivity, with a magnitude larger than the uncertainty in the pseudo-T-odd contribution from final state interactions.

Our discussion of this analysis is organized as follows. In Sec.~\ref{EDMsystemssection} we review the basic interactions and the expression for the EDM of hadronic, atomic and molecular systems. In Sec.~\ref{mLRSMsection} we discuss the relevant interactions within the mLRSM. In Sec.~\ref{globalsection}, we present our results for a global fit using the neutron, Xenon (Xe), Radium (Ra), and Thallium fluoride (TlF) systems. We discuss our results in Sec.~\ref{discussionsection}. Then, 
Sec.~\ref{Dsection} gives our analysis of the interplay between T-reversal violation in $\beta$-decays and the EDM bounds implied by our global fit. Finally in Sec.~\ref{conclusions} we present our conclusions. 	
\section{EDM of hadronic and atomic systems } \label{EDMsystemssection}

In this section we review and summarize the expressions needed for computing the EDM of hadronic, atomic and molecular systems. We focus on the EDMs of diamagnetic systems for the following reasons. The parameter that governs the magnitude of the CP-violating couplings is $\kappa_{LR}$ --defined below -- which arises from mixing between left- and right-handed gauge bosons, $W_{L,R}$ and generates a dimension six four quark operator. The resulting, leading contribution to paramagnetic systems is given by the electron EDM ($d_e$). In the mLRSM, $d_e$ arises at one-loop order, introducing a the general one-loop suppression factor $1/16\pi^2$~\cite{Valle:1983nx,Nieves:1986uk}. Moreover, $d_e$ depends on the neutrino Dirac mass parameter $M_D$ that does not enter the diamagnetic EDMs at an appreciable level. Consequently, one may treat $d_e$ as providing a separate constraint on $M_D$, while the diamagnetic systems constrain ${\bar\theta}$ and the left-right operator defined below.   Previous $d_e$ bounds  require  $M_D \lesssim (10^{-2}-1)$ MeV~\cite{Tello:2012qda}. 
In addition to the above theoretical arguments,  the new experimental limit  $d_e < 10^{-29}$ e.cm~\cite{Baron:2013eja} implies a corresponding $d_e$ contribution to the $^{199}$Hg EDM to be  $\lesssim 10^{-31}$ e.cm , below the current sensitivity $d_{A}\left({ }^{199} \mathrm{Hg}\right)=(2.20 \pm 2.75(\text { stat }) \pm 1.48(\mathrm{sys})) \times 10^{-30} e\cdot \mathrm{cm}$~\cite{PhysRevLett.119.119901}. The individual quark contributions to the nucleon EDMs are suppressed due to the small Yukawa coupling of the up and down quarks~\cite{Engel:2013lsa}. The  resulting contributions yield a weaker sensitivity to the parameters of the mLRSM than the effects considered below. Finally, the three-gluon Weinberg operator gives subleading contributions to the EDMs for light quarks -- as discussed in Ref.~\cite{Bertolini:2019out}.

For diamagnetic systems such as the neutron and $^{199}$Hg atom, the relevant contributions are the EDM and chromo-EDM of the quarks, the CPV three-gluon operator, and various four fermion operators. The quark (chromo-)EDM and three-gluon operators also arise at one-and two-loop orders~\cite{Xu:2009nt}, respectively. The former carry factors of the light quark Yukawa couplings, introducing an additional source of suppression. In contrast, $W_L-W_R$ mixing can also yield a CPV four-quark operator via tree-level gauge boson exchange as indicated in Fig.~\ref{diagramLR}. The corresponding Wilson coefficient is also proportional to $\kappa_{LR}$ but carries neither a loop nor Yukawa suppression factors. Thus, one expects it to yield the leading contribution to diamagnetic EDMs.
At the hadronic level, this dimension six operator generates a ``short distance" contribution to the nucleon EDM and the leading-order (LO) chiral contribution coming from one-loop diagrams involving the CP-violating interaction between the pions and the nucleons. Finally, the dimension six operator also contributes to the atomic and molecular EDMs via the nuclear Schiff moment -- see Ref.~\cite{Engel:2013lsa} for details.

 The EDM of diamagnetic atomic or molecular system is given by~\cite{Chupp:2014gka}: 
\begin{equation}
d_{A}=\sum_{N=p, n} \rho_{Z}^{N} d_{N}+\kappa_{S} S-\left[k_{T}^{(0)} C_{T}^{(0)}+k_{T}^{(1)} C_{T}^{(1)}\right],
\end{equation}
where the parameters $\rho_{Z}^{N}$ give the sensitivity of the EDMs to the individual nucleons. The parameter $\kappa_{S}$ measures the sensitivity of the atomic or molecular system to the nuclear Schiff moment $S$. For the mLRSM, one has
\begin{equation}\label{Schiff}
S\simeq\frac{m_{N} g_{A}}{F_{\pi}}\left[a_{0} \bar{g}_{\pi}^{(0)}+a_{1} \bar{g}_{\pi}^{(1)}\right],
\end{equation}
where the couplings $\bar{g}_{\pi}^{(i)}$ parametrize the T-violating, P-violating pion-nucleon interaction in the chiral Lagrangian
\begin{equation}
\mathcal{L}_{\chi}^{LR}=\bar{N}\left[\bar{g}_{\pi}^{(0)} \vec{\tau} \cdot \vec{\pi}+\bar{g}_{\pi}^{(1)} \pi^{0}\right] N \ \ \ .
\label{chiral_lagrangian}
\end{equation}
It is useful to express the $\bar{g}_{\pi}^{(i)}$ in terms of the Wilson coefficient of the four-quark operator associated with Fig.~\ref{diagramLR}
\begin{equation}
\bar{g}_{\pi}^{(i)} = \lambda_i \bar{\theta} + \gamma_i^{\varphi ud} \frac{v^2}{\Lambda^2}\operatorname{Im} (C_{\varphi ud}),\, i = 0,1,\, ,\label{gis}
\end{equation}
where the dimension-six effective Lagrangian contains the term
\begin{equation}\label{dim6}
-i \frac{\operatorname{Im} C_{\varphi u d}}{\Lambda^{2}}\left[\bar{d}_{L} \gamma^{\mu} u_{L} \bar{u}_{R} \gamma_{\mu} d_{R}-\bar{u}_{L} \gamma^{\mu} d_{L} \bar{d}_{R} \gamma_{\mu} u_{R}\right]\, .
\end{equation}
We discuss the range of numerical values for the $\lambda_i$ and $\gamma_i$ below.

Finally, $C_T^{0}$, $C_T^{1}$ are related to the nucleon-electron interaction of the form:
\begin{equation}\label{semileptonic}
\mathcal{L}_{e N}^{\mathrm{NSD}}=\frac{8 G_{F}}{\sqrt{2}} \bar{e} \sigma_{\mu \nu} e v^{\nu} \bar{N}\left[C_{T}^{(0)}+C_{T}^{(1)} \tau_{3}\right] S^{\mu} N+\cdots.
\end{equation}
where $S^{\mu}$ denotes the nucleon spin, which in the nucleon rest frame takes the value $(0,\vec{\sigma}/2)$ and $\sigma_i$ are the Pauli matrices. This interaction enters in the atomic EDM due to the interaction of the closed electron shells with the hyperfine interaction polarizing these shells along the nuclear spin~\cite{Khriplovich:1997ga}. The constants $C_{T}^{(0)}$ and $C_{T}^{(1)}$ parametrize the nuclear-spin-dependent electron-nucleon interaction. Notice that all the semileptonic operators at dimension six can only be obtained by integrating out neutral currents at the tree level~\cite{Engel:2013lsa}. The diagrams involving the SM Higgs exchange (on top of being suppressed by small lepton and light quark Yukawa couplings) cannot give the Lorentz structure show in Eq.~\eqref{semileptonic}. The same reasoning applies to the exchange of hypothetical new heavy scalar particles and the right-handed $Z$-boson of the mLRSM. 

\section{The minimal left-right symmetric model}\label{mLRSMsection}
\begin{figure}
 \centering
 \includegraphics[width=.6\columnwidth]{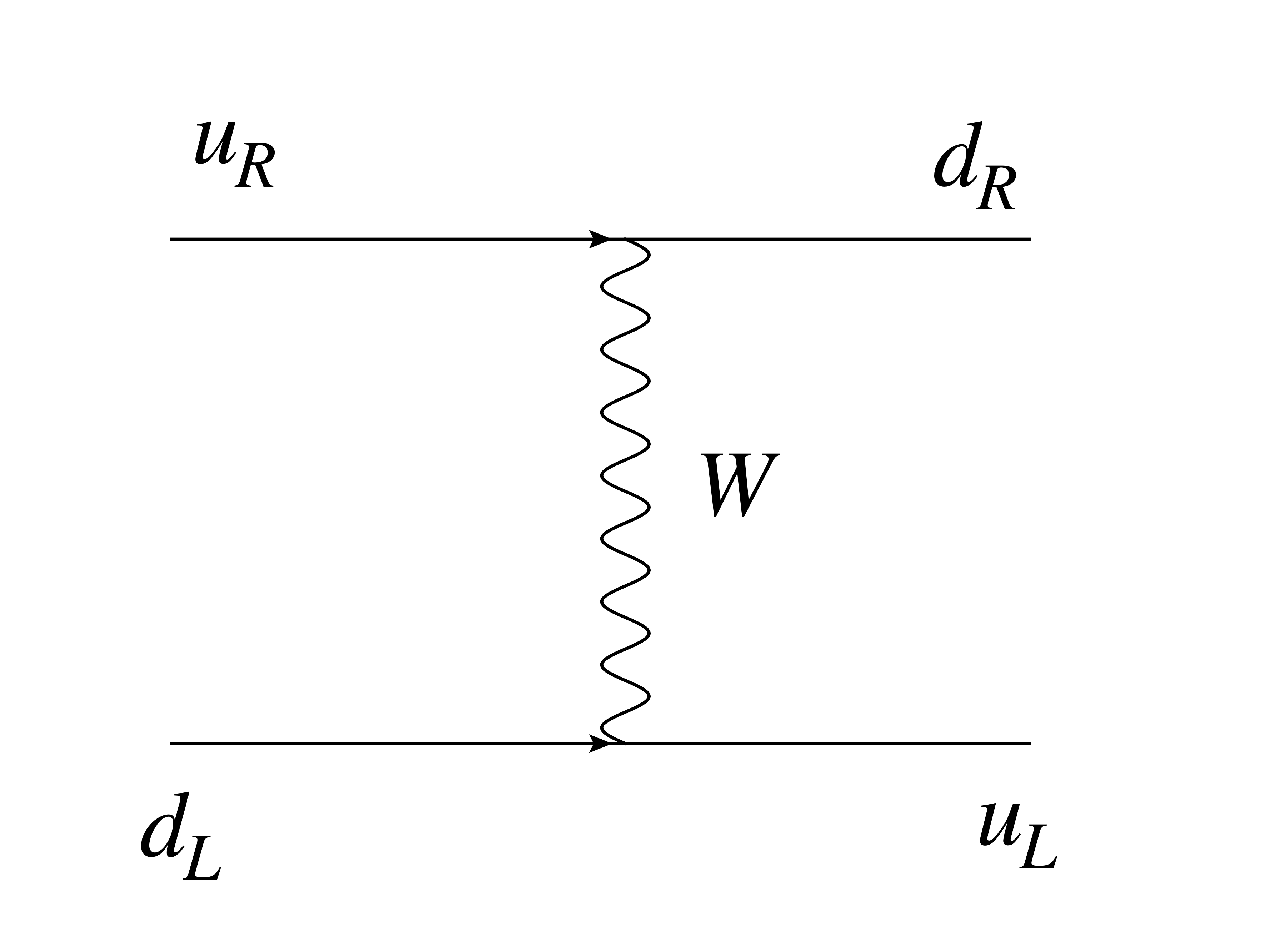}
 \caption{ Leading tree-level diagram contributing to the neutron EDM in the mLRSM. }
 \label{diagramLR}
\end{figure}
The mLRSM~\cite{SENJANOVIC1979334,Mohapatra:1979ia,Mohapatra:1980yp} extends the SM gauge group to $SU(3)\times SU(2)_R\times SU(2)_L \times U(1)_{B-L}$, where $B$ and $L$ denote the SM abelian baryon and lepton quantum numbers. In this model the $SU(2)_R$ gauge group is broken once the triplet field 
 \begin{equation}
 \Delta_{R}=\left(\begin{array}{cc}\Delta^{+} / \sqrt{2} & \Delta^{++} \\ \Delta^{0} & -\Delta^{+} / \sqrt{2}\end{array}\right)_{R}
 \end{equation}
takes a vacuum expectation value (VEV) $v_R$ along with its neutral component, breaking the above gauge group down to the SM gauge group. At the next step of symmetry breaking, the SM gauge group is broken down to $U(1)_{QED}$ once the bidoublet field
\begin{equation}
 \Phi=\left(\begin{array}{cc}\phi_{1}^{0} & \phi_{2}^{+} \\ \phi_{1}^{-} & \phi_{2}^{0}\end{array}\right)
\end{equation}
takes a VEV along its neutral components $\langle \Phi \rangle = \text{diag}\, \{v_1, v_2 e^{i \alpha}\}$. Notice the appearance of the phase $\alpha$. This is the so-called spontanous CP phase, which gives an EDM contribution, as we discuss below.

In the mLRSM one has, in addition to the $\bar{\theta}$ contribution to the EDMs, a contribution from the tree-level diagram shown in Fig.\ref{diagramLR}, which is due to the interaction 
 \begin{align} 
 \mathcal{L}^{W}=&-\frac{g}{\sqrt{2}} \bar{u}_{L i} \gamma^{\mu}V^L_{ud}\left(\cos \xi W_{1 \mu}^{+}-\sin \xi e^{-i \alpha} W_{2 \mu}^{+}\right) d_{L i} \nonumber \\ &-\frac{g}{\sqrt{2}} \bar{u}_{R i} \gamma^{\mu}V^R_{ud}\left(\sin \xi e^{i \alpha} W_{1 \mu}^{+}+\cos \xi W_{2 \mu}^{+}\right) d_{R i}\nonumber \\ 
 &
 + \text { h.c. }\,,
 \end{align}
where $W_1$ is light mass eigenstate, which is mostly the SM $W$ boson, $W_2$ is the heavy mass eigenstate, and $V^{L,R}$ are the CKM matrix and its right-handed analog, respectively. By the exchange of one $W$ boson shown in Fig.~\ref{diagramLR}, this interaction generates a four-quark operator proportional to $\kappa_{LR}$, with the Wilson coefficient suppressed by the mixing between the SM $W$ and the $W_R$ bosons. The resulting CPV effective Lagrangian involving light quarks and gluons, valid at the weak scale, is given by
\begin{align}\label{dim6}
\mathcal{L}_{CPV} =&-\frac{g_{3}^{2}}{16 \pi^{2}} \bar{\theta} \operatorname{Tr}\left(G^{\mu \nu} \tilde{G}_{\mu \nu}\right)- \nonumber \\ &i \frac{4G_F}{\sqrt{2}}\kappa_{LR}\left( \bar{u}_R\gamma_{\mu}d_R\, \bar{u}_L\gamma_{\mu}d_L + h.c. \right), 
\end{align}
where, 
\begin{equation}
\kappa_{LR}= \sin \xi Im \left( V_{ud}^LV_{ud}^{R*}e^{-i\alpha}	\right)\ \ \ , 
\end{equation}
\begin{equation}
 \tan\xi = - \frac{v_1v_2}{v_R^2}\simeq - \frac{M_W^2}{M_{W_R}^2}\sin2\beta\ \ \ ,
\end{equation}
with $\tan\beta \equiv v_2/v_1$, 
 $g_3$ is the strong coupling constant, $G_{\mu \nu}$ is the gluon field strength and $\tilde{G}_{\mu \nu}=\epsilon_{\mu \nu \alpha \beta} G^{\alpha \beta} / 2$.  

 As emphasized above, $\bar\theta$ and $\kappa_{LR}$ are independent parameters when $\mathcal{C}$ is taken as the LR symmetry. Both terms in Eq.~(\ref{dim6}) induce the neutron EDM through (a) \lq\lq long-distance\rq\rq\, chiral loops arising from the pion nucleon interaction in Eq.~\eqref{chiral_lagrangian}; and (b) \lq\lq short distance\rq\rq\, effects encoded in the finite parts of the low energy constants of the associated chiral effective field theory. The same operator also generates a contribution to atomic and molecular EDMs via the nuclear Schiff moment $S$ in Eq~\eqref{Schiff}. Owing to their different chiral transformation properties, the two interactions in Eq.~(\ref{dim6}) generate leading contributions to the two different $\bar{g}_{\pi}^{(i)}$, {\em viz} $\bar{g}_{\pi}^{(0}\sim {\bar\theta}$ and $\bar{g}_{\pi}^{(1)}\sim \kappa_{LR}$. 
 
We take into account the RGE from the electroweak scale down to the hadronic scale using the results of Refs.~\cite{deVries:2012ab,Dekens:2013zca,Cirigliano:2016yhc}. In addition to the RGE effects, the ``direct" part contributing to the finite part of the low energy constant was studied in detail in Ref.~\cite{Cirigliano:2016yhc}, where it 
induces a roughly $50\%$ uncertainty in the $\kappa_{LR}$ dependence of the neutron EDM. The net effect of these two issues is reflected in the estimated range for the coefficient $\gamma_1^{\varphi ud}$  reported in Table 7 of Ref.~\cite{Engel:2013lsa}. In this work we update the range of the $\gamma_{1}^{\varphi ud}$ parameter reported in Ref.~~\cite{Engel:2013lsa}\footnote{ The coefficient $\gamma_{1}^{\varphi ud}$ of this paper is denoted as $\gamma_{(1)}^{\varphi ud}$ in Ref.~\cite{Engel:2013lsa}. }. The previous range $\gamma_1^{\varphi ud}\in (5-150)\times10^{-7}$ of Ref.~\cite{Engel:2013lsa}, once the RGE effects and the short distance effect is taken into account changes to $\gamma_1^{\varphi ud}\in (254-552)\times10^{-7}$. We have estimated this new range by adding in quadrature the uncertainties reported in Eqs. (31) and (32) of Ref.~\cite{Cirigliano:2016yhc}. The short distance contribution to the nucleon EDM was studied in detail in Ref.~\cite{Engel:2013lsa} and we use the range reported for the parameter $\beta_n^{\varphi ud}$ in Table 7 of Ref.~\cite{Engel:2013lsa}. 

\begin{figure*}
	\centerline{%
\hspace{-0.9em}
		\includegraphics[width=.68\columnwidth]{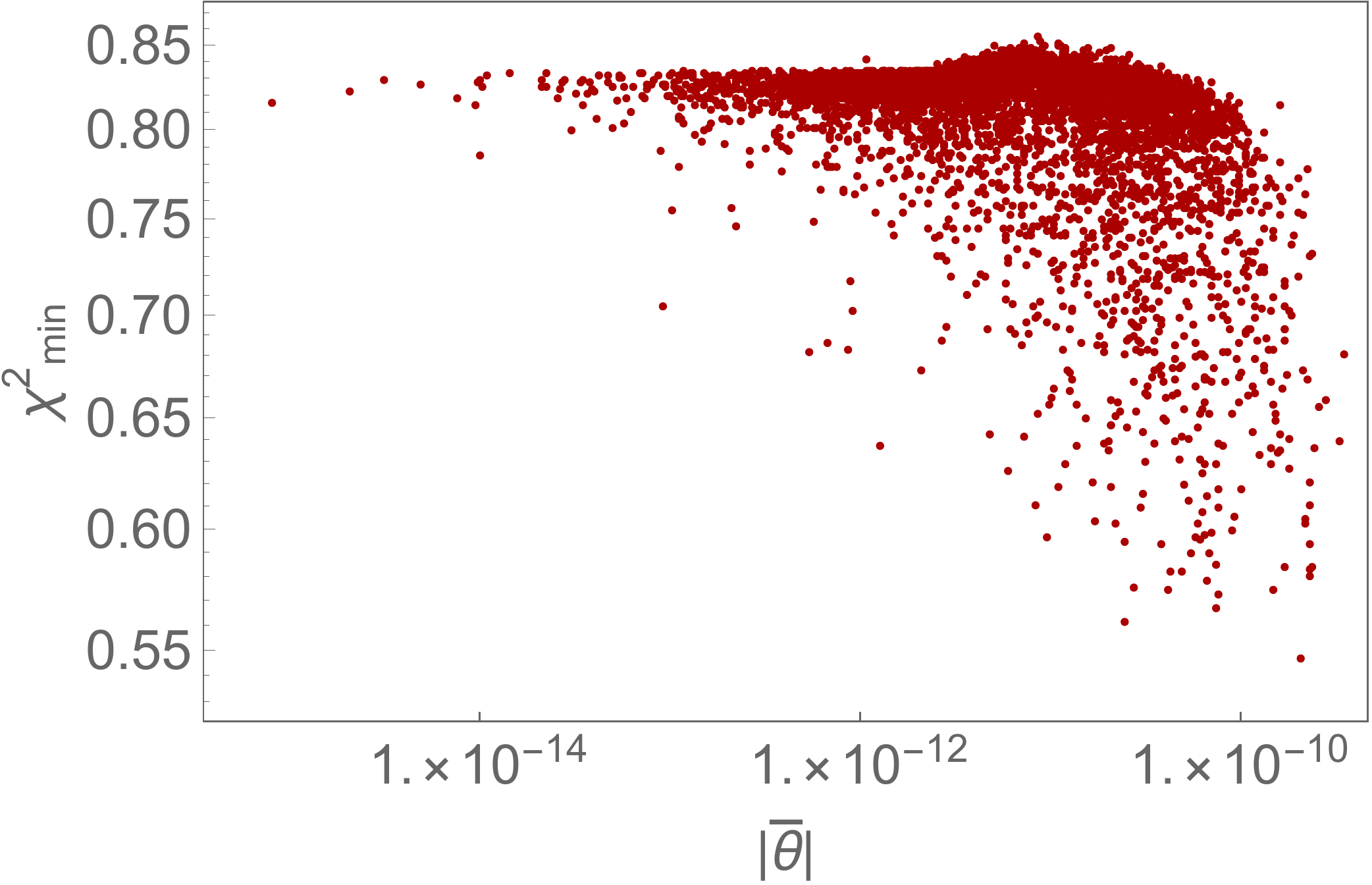}~~~~~~~~
\hspace{-2.06em}
		\includegraphics[width=.68\columnwidth]{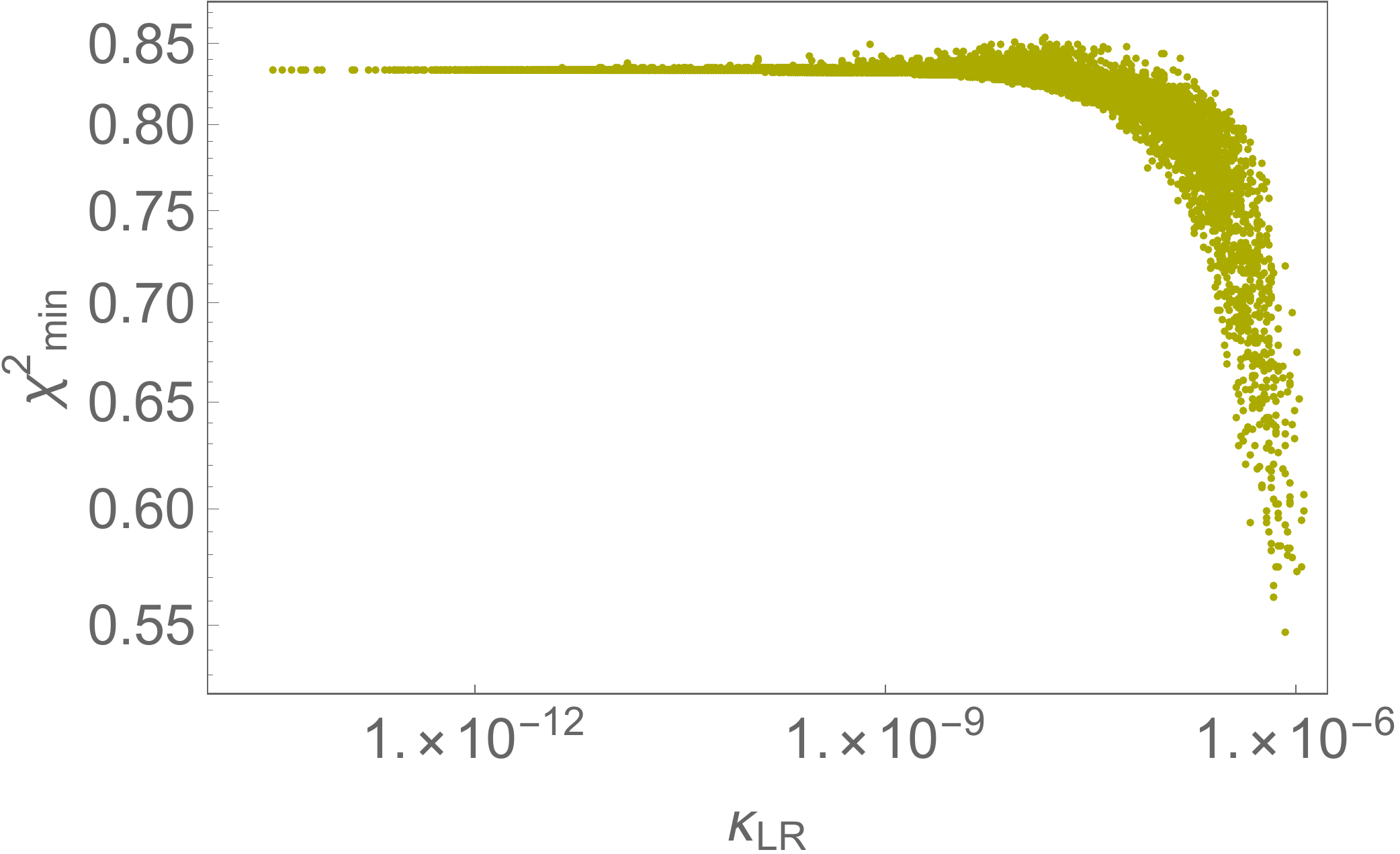}%
\hspace{-0.53em}
		 \includegraphics[width=.68\columnwidth]{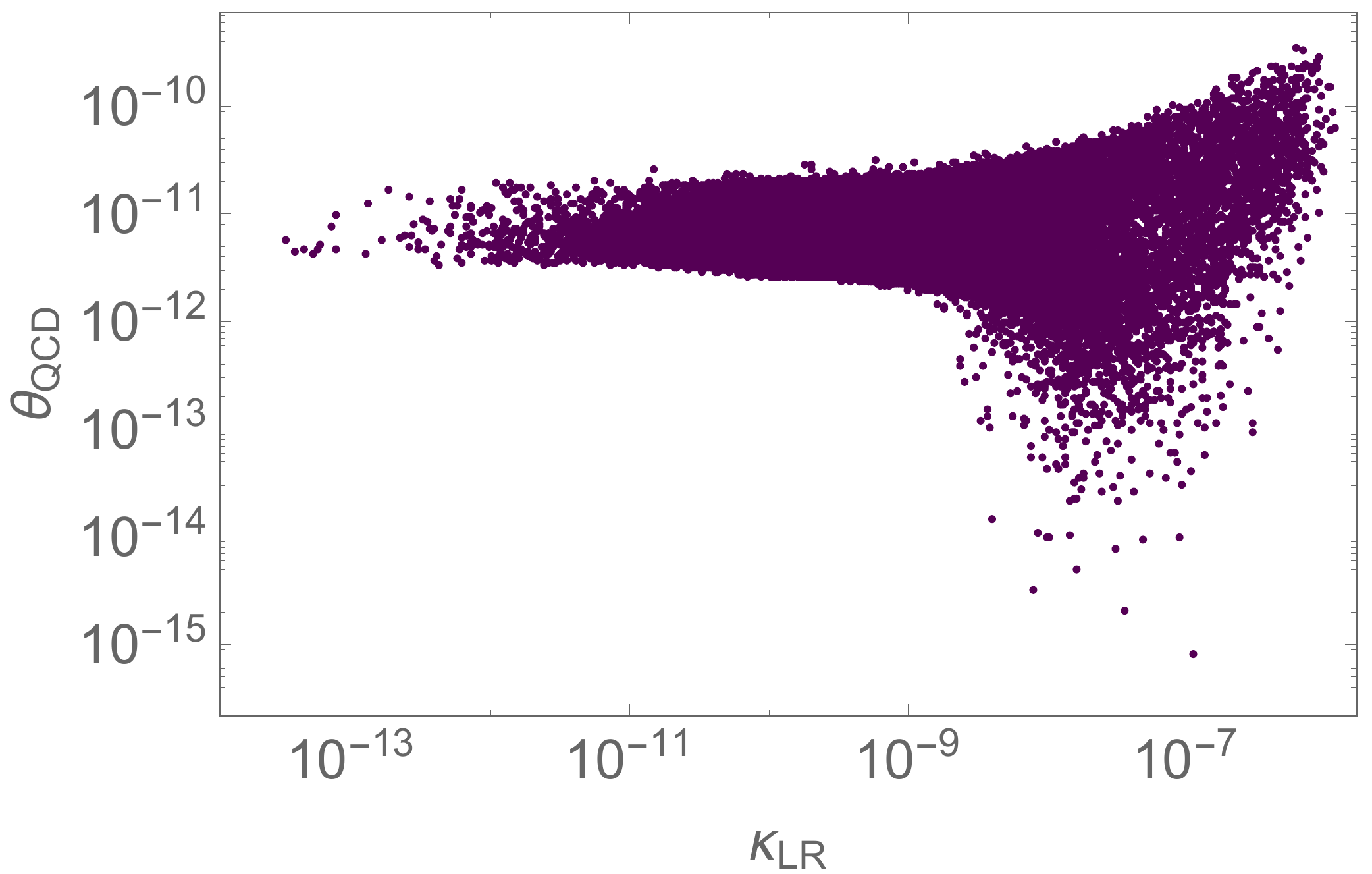}%
	}%
	\vspace*{-2ex}
	\caption{ Left. Values of $\chi^2_{min}$ as a function of the $\bar{\theta}$ contribution. Center. Values of $\chi^2_{min}$ as a function of the LR contribution. Right. Values of $\bar{\theta}$ as a function of the LR contribution}\label{CHI2VSTHETA}
\end{figure*}
\section{Global analysis using EDMs of nucleons, atoms and molecules } \label{globalsection}

In this section, we perform a global fit using the most recent results for the EDMs of the neutron, atomic and molecular systems Hg, Xe, Ra, and TlF taken from Ref.~\cite{Chupp:2017rkp}. We use the value of the hadronic and nuclear parameters given in Ref.~\cite{Chupp:2014gka}~\footnote{Recently, the values of the nuclear parameters for $\mathrm{Hg}$ and $\mathrm{Xe}$ were calculated in Ref.~\cite{Yanase:2020agg} and fall between the ranges reported in Ref.~\cite{Chupp:2014gka}. } We take the centroid for the experimental values of the EDM from Ref.~\cite{Chupp:2017rkp} and references therein. The most recent neutron EDM experimental limit was taken from Ref.~\cite{Abel:2020gbr}. 

In order to find the sensitivity of the nucleon, atomic and molecular EDM systems to the $\bar{\theta}$ parameter and the mLRSM contribution, we perform a $\chi^2$ fit (with 2 d.o.f.) using the function: 
\begin{equation}
\chi^2 = \sum_{i=1}^{N} \frac{\left[(d_i)_{exp}-(d_i)_{th}\right]^2}{\sigma_i^2},
\label{chi-square}
\end{equation}
where $N$ is the number of the EDM systems, $(d_i)_{exp}$ and $(d_i)_{th}$ denotes the experimental centroids and the theoretical values for the EDMs and $\sigma_i$ denotes the experimental error of the EDM for the system $i=n$, Hg, Xe, Ra, TlF. To take into account the theoretical uncertainties of the hadronic, nuclear, and atomic parameters, we perform a range fit~\cite{Charles:2004jd}, which we explain in detail in what follows: 
\begin{itemize}
\item We vary the parameters $a_0,a_1$ and between the best theoretical ranges reported in Tab. VI of Ref.~\cite{Chupp:2014gka}, except for the TlF, since the ranges for these parameters are not reported in Ref.~\cite{Chupp:2014gka}. For the parameter $\kappa_S$, we vary it only for Ra in the range reported in Tab. VI of Ref.~\cite{Chupp:2014gka}. For Mercury, Xenon, and TlF  we use the fixed values of $\kappa_S$ reported in Tab. VI of Ref.~\cite{Chupp:2014gka}. 

\item The hadronic uncertainty is taken into account by varying the parameter $\alpha_n$, $\beta_n^{\varphi ud}$ in Eq.~\eqref{dn} and  $\lambda_0$, $\lambda_1$ and $\gamma_1^{\varphi ud}$ in Eq.~\eqref{gis} within the range reported in Table 7 of Ref.~\cite{Engel:2013lsa}. The parameter $\gamma_0^{\varphi ud}=0$, since the four quark operator in Eq.~\eqref{dim6} breaks isospin symmetry, and hence it cannot contribute to the coupling $\bar{g}_{\pi}^{(0)}$ in Eq.\eqref{chiral_lagrangian}, which conserves isospin. The only difference is that we update the range for the parameter $\gamma_1^{\varphi ud}$ and use the new one discussed in Sec.~\ref{mLRSMsection}.  Finally, using the more recent results of Ref.~\cite{deVries:2015una}, we also update the range for the parameter $\lambda_{0}$. The updated ranges are summarized in Tab.~\ref{tab:update}. 

\begin{table} 
\centering 
\begin{tabular}{l c c } 
\toprule 
\midrule 
\textbf{Hadronic parameter} & Updated range\\ 
\midrule 
$\gamma_1^{\varphi ud}$& $(254-552)\times10^{-7 }$ \\ 
$\lambda_0$& $0.013-0.018$ \\ 
\midrule 
\bottomrule 
\end{tabular}
\caption{ Updated ranges for two of the parameters reported in Table 7 of Ref.~\cite{Engel:2013lsa}. We used the results of Ref.~\cite{Cirigliano:2016yhc} and Ref.~\cite{deVries:2015una}. The coefficients $\gamma_i^{\varphi ud} \,(\lambda_i)$ in this paper are denoted as $\gamma_{(i)}^{\varphi ud}, \,(\lambda_{(i)})$ in Ref.~\cite{Engel:2013lsa}, for $i = 0,1$. } 
\label{tab:update} 
\end{table}
\begin{figure}
 \centering
 \includegraphics[width=1\columnwidth]{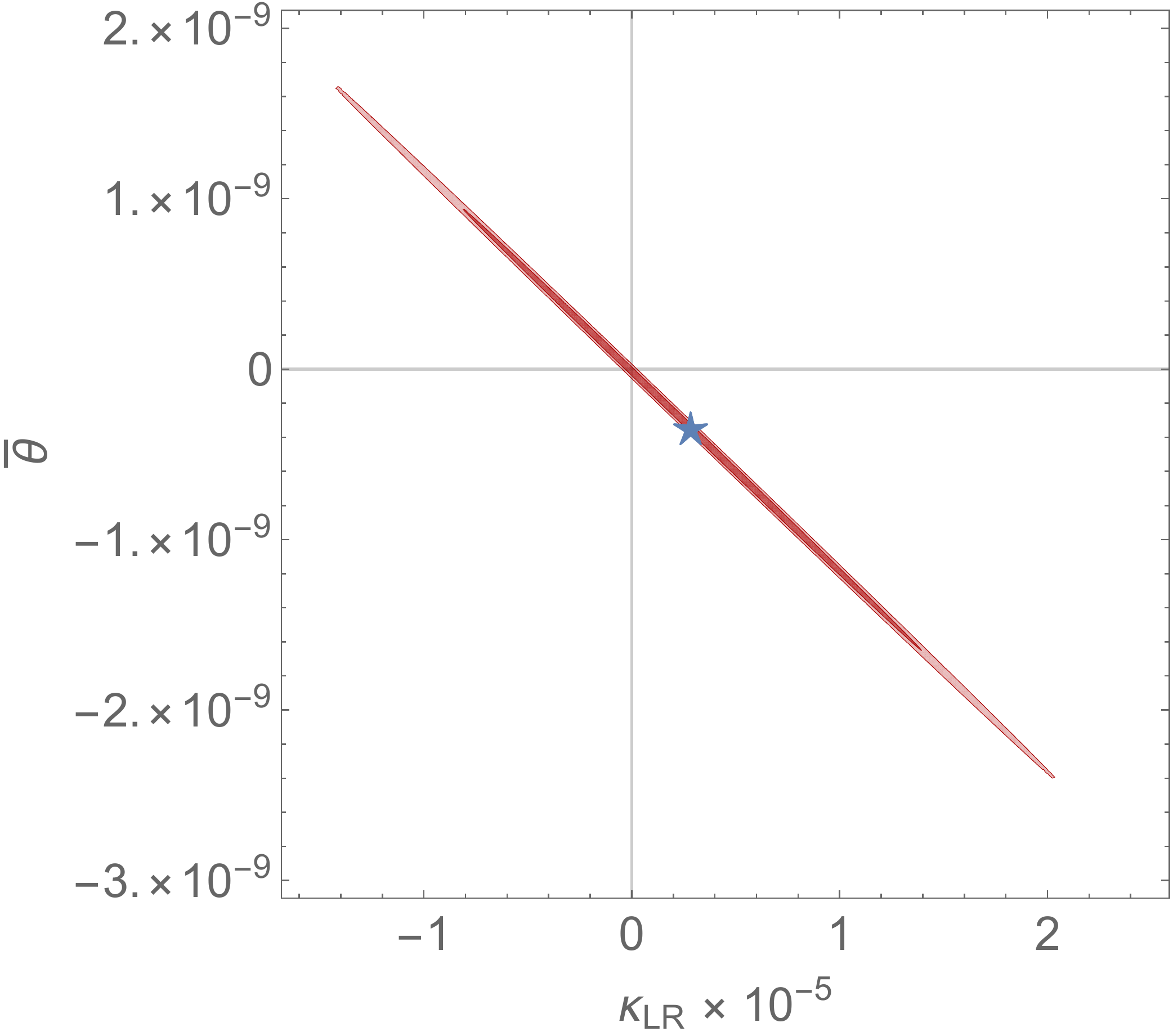}
 \caption{Ranges represent the $68\%$ and $95\%$ confidence level interval resulting from the best fit for the diamagnetic systems including the neutron EDM. The blue star indicates the position of the best fit point. }
 \label{CHI2}
\end{figure}

\item For each point in the space spanned by $a_0,a_1$, $\kappa_s$, $\alpha_n$, $\beta_n^{\varphi ud}$, $\lambda_0$, $\lambda_1$ and $\gamma_1^{\varphi ud}$ 
 we minimize the $\chi^2$ in Eq.~\eqref{chi-square} with respect to the $\bar{\theta}$ and $\kappa_{LR}$. 

\item From all possible values of the $\chi^2_{min}$ function, we choose those values that give the most conservative bound. By the most conservative bound we mean that choosing any other value for $a_0,a_1$ and $\kappa_S$, $\alpha_n$, $\lambda_0$, $\lambda_1$ and $\gamma_1^{\varphi ud}$ within the ranges reported in Ref.~\cite{Chupp:2014gka} and Tab.~\ref{tab:update} gives  a stronger bound. It turns out the values of $\chi^2_{min}$ giving the more conservative bound are strongly correlated with the smallest values of $\chi^2_{min}$ as can be seen from Fig.~\eqref{CHI2VSTHETA}. Finally, notice from Fig.~\eqref{CHI2VSTHETA} (right) that the largest values of $\bar{\theta}$  are also correlated with the largest values for $\kappa_{LR}$. 

\item Once the values of the hadronic parameters $\lambda_0$, $\lambda_1, \gamma_1^{\varphi ud}$, $\alpha_n$ and $\beta_n^{\varphi ud}$ , the nuclear parameters $a_0$, $a_1$ and the atomic parameter $\kappa_s$, which give the weakest bound are found, we determine the $68\%$ and $95\%$ C.L. intervals as shown in Fig.~\eqref{CHI2}. In Fig.~\eqref{CHI2}, we plot the 68$\%$ and 95$\%$ confidence level interval in  the $(\bar{\theta}$, $\kappa_{LR})$ plane. The $\bar{\theta}$ and $\kappa_{LR}$  parameters are bounded to be inside the colored regions. The blue star denotes the location of the best-fit value. This value is closer to the limit $\bar{\theta}\lesssim 4\times10^{-10}$ ($\kappa_{LR}\lesssim 3\times10^{-6}$) derived when assuming the dominance of the $\bar{\theta}$ term (or the $\kappa_{LR}$ contribution). As can be seen, these sole-source bounds get relaxed by roughly one order of magnitude when doing the global analysis. 
\end{itemize}
 For completeness, we show in Tab.~\ref{tab:atomic} the best fit values values for the  atomic and nuclearparameters used in our global fit. In Tab.~\ref{tab:hadro}  we give the best fit values for the hadronic parameter $\alpha_n$, $\beta_n$, $\lambda_0$, $\lambda_1$ and $\gamma_1^{\varphi ud}$. 

\begin{table} 
\centering 
\begin{tabular}{l c c c } 
\toprule 
\midrule 
& \multicolumn{3}{c}{\textbf{Atomic and nuclear parameters}} \\ 
\cmidrule(l){2-4} 
\textbf{EDM System} & $\kappa_S(\text{fm}^{-2})$ & $a_0$ & $a_1$ \\ 
\midrule 
Mercury ($\mathrm{Hg}$) & $-2.8\times10^{-4}$ & $0.022$ & $0.0029$ \\ 
 Xenon ($\mathrm{Xe}$)& $2.7\times10^{-5}$ & -0.036 & -0.024 \\ 
Radium ($\mathrm{Ra}$) & $-7.6\times10^{-4}$ & -3.45 & 5.1 \\ 
 Tellure Fluoride TlF  & -0.74 & -0.012 & 0.16 \\ 
\midrule 
\bottomrule 
\end{tabular}
\caption{ Best fit values of the Schiff moments and the dependence of the Schiff moment on $ \bar{g}_{\pi}^{(0)}$ and $ \bar{g}_{\pi}^{(1)}$ that give the more conservative bounds in the allowed ranges reported in Ref.~\cite{Chupp:2014gka}. These are the values used to obtained the confidence level intervals shown in Fig.~\eqref{CHI2}. } 
\label{tab:atomic} 
\end{table}
\begin{table} 
\centering 
\begin{tabular}{l c c } 
\toprule 
\midrule 
\textbf{Hadronic parameter} & Best fit value \\ 
\midrule 
$\alpha_n[e\cdot\text{fm}]$& $0.5\times10^{-3 }$ \\ 
$\beta_n^{\varphi ud}[e\cdot\text{fm}]$& $8.4\times 10^{-8}$ \\ 
 $\lambda_0$& $0.017$ \\ 
$\lambda_1$ & $2.7\times10^{-4}$ \\ 
$ \gamma_1^{\varphi ud}$& $311\times10^{-7}$ \\ 
\midrule 
\bottomrule 
\end{tabular}
\caption{ Best fit values of the hadronic parameters that gives the more conservative bound for their allowed ranges reported in Table 7 of Ref.~\cite{Engel:2013lsa}. The range of the parameter $\gamma_1$ has been updated to include the RGE effects and the short distance uncertainty reported in Ref.~\cite{Cirigliano:2016yhc} } 
\label{tab:hadro} 
\end{table}

 In Fig.~\eqref{CHI2VSa1Hg} and Fig.~\eqref{CHI2VSa1Ra} of the Appendix, we show the value of $\chi^2_{min}$ as a function of $a_1(\mathrm{Hg})$ and $a_1(\mathrm{Ra})$. Notice that the best fit prefer small and positive values of the parameter $a_1(\mathrm{Hg})$, whereas there is a preference for the smallest value of $a_1(\mathrm{Ra})$. For the other parameter, we do not find this correlation and hence these are not interesting to report. 

\section{The Strong CP problem in the minimal LR symmetric model} \label{discussionsection}

The strong CP problem within the mLRSM has been recently discussed Refs.~\cite{Maiezza:2010ic,Maiezza:2014ala,Senjanovic:2015yea,Senjanovic:2020int}. For the sake of completeness and to emphasize the usefulness of our fit we highlight the main points here. As a concrete example take the expression of the neutron EDM
\begin{align}\label{nEDM}
 d_n \simeq \alpha_n\,\bar{\theta}+ \beta_n^{\varphi ud} \,\left(\frac{v^2}{\Lambda^2}\right)\,Im \left(C_{\varphi ud}	\right), 
\end{align}
where
\begin{equation}
\left(\frac{v^2}{\Lambda^2}\right) Im \left(C_{\varphi ud}\right) = \kappa_{LR}
\end{equation}
and $\alpha_n $ and $\beta_n^{\varphi ud}$ are given in Tab.~\ref{tab:hadro}. In what follows we discuss the situation for both $\mathcal{P}$ and $\mathcal{C}$ as the LR symmetry. 

\paragraph{Parity as the left-right symmetry.}
For $\mathcal{P}$, one has $V_{ud}^{L} = V_{ud}^{R} + \mathcal{O}(\tan 2\beta\sin\alpha)$~\cite{Senjanovic:2014pva,Senjanovic:2015yea}, and at leading order in $v_2/v_1$ we have the following relation 
\begin{align}\label{EDMP}
 d_n \simeq \alpha_n\,\bar{\theta} - \beta_n^{\varphi ud} \,\left(\frac{M_W^2}{M_{W_R}^2}\right)\, |V_{ud}^L|^2\sin2\beta \sin\alpha +\mathcal{O}(v_2^2/v_1^2)
\end{align}

 Under the hypothesis that $\mathcal{P}$ is an exact symmetry, $\theta_0=0$. In this case $\bar{\theta}\simeq \frac{m_t}{2m_b}\tan 2\beta\sin\alpha$~\cite{Maiezza:2014ala}, where $m_b$ and $m_t$ are bottom and top quark masses, respectively. Invoking an explicit breaking of $\mathcal{P}$ in the strong sector alone, such that $\theta_0$ cancels the $ \frac{m_t}{2m_b}\tan 2\beta\sin\alpha$ contribution to $\bar{\theta}$ would not work. As discussed in Ref.~\cite{Maiezza:2014ala}, by doing an anomalous chiral rotation, a tiny explicit breaking in the strong sector would induce a large amount of explicit breaking in the Yukawa sector, and the relation $V^L\simeq V^R$ would be lost (this  has been recently discussed  in  Ref.~\cite{Maiezza:2020fcv}).

 By requiring the $\bar{\theta}$ parameter to be smaller than its experimental limit $\theta_{exp}\sim 10^{-9}$ (see Fig.~\ref{CHI2}) one gets $\tan 2\beta\sin\alpha\leq \frac{2m_b}{m_t} \theta_{exp}\sim 10^{-12}$, which makes all the phases in $V_R$ effectively zero -- with exception of the Dirac phase $\delta_{CKM}$, which gives a negligible contribution to the neutron EDM $\sim10^{-32}\,e\cdot$cm~\cite{Seng:2014lea}. The parameter $\tan2\beta\sin\alpha$ gauging the $\bar{\theta}$ contribution is essentially the same one entering in the new physics corrections shown in the second term of Eq.\eqref{EDMP}. Therefore, since $\beta_n^{\varphi ud}\ll \alpha_n $ one can write the neutron EDM as 
\begin{align}\label{EDMP_approx}
 d_n \simeq \alpha_n\,\bar{\theta}. 
\end{align}
Whereby, if $\mathcal{P}$ is an exact symmetry, one gets the prediction that all the EDM should be dominated by the $\bar{\theta}$ term and this would give precise patterns for the ratios of different EDMs. Note that as argued in Ref.~\cite{deVries:2018mgf} in the context of the SM effective field theory for dim-6 operators, if the strong CP problem is solved by ultraviolet physics, then the low energy EDMs should be dominated by the $\bar{\theta}$ term. Similar considerations of using EDM ratios to single out the underlying mechanism of CP violation have also been considered in Refs.~\cite{Lebedev:2004va,Bsaisou:2014oka,Dekens:2014jka}. 

\paragraph{Charge conjugation as the left-right symmetry.}
For $\mathcal{C}$ the parameter $\bar{\theta}= \theta_0+\text{arg det}(M_uM_d) $ with $\theta_0\neq 0$. This means that both the $\bar	{\theta}$ and the new physics contributions in Eq.~\eqref{nEDM} are independent from each other. In this case one cannot get any limit using one EDM, since one cannot invoke the idea of parity invariance to relate the first with the second term in Eq.~\eqref{nEDM}. In Fig.~\ref{CHI2}, we show the result of the global fit using several EDM systems. We found that the typical limit $\bar{\theta}\sim1.5\times10^{-10}$~\cite{Dragos:2019oxn} one gets when using one EDM system gets relaxed to $|\bar{\theta}|\simeq 2.4\times10^{-9}$ with a 95$\%$ of confidence level. The same reasoning applies to the $\kappa_{LR}$ contribution. Unlike the case of $\mathcal{P}$, for $\mathcal{C}$ as the LR symmetry, the ratio between different EDM do not are necessarily dominated by the $\bar{\theta}$ term contribution. 

\section{Interplay with time-reversal symmetry violation in $\beta$-decays} \label{Dsection}
In this section, we discuss the implication of our global fit to the so called ``D" coefficient in polarized $\beta$-decays~\cite{Jackson:1957zz}. The neutron differential decay rate contains the triple-vector product $d\Gamma/d\Omega \supset D\, \langle \mathbf{\vec{J}}\rangle\cdot \vec{p}_e\times \vec{p}_{\nu}$, where $\langle \mathbf{\vec{J}}\rangle$ is the nuclear polarization of the neutron, $\vec{p}_e$ and $\vec{p}_{\nu}$ are the three-momentum of the outgoing electron and neutrino, respectively. This coefficient can be decomposed as $D = D_f + D_t$, where $D_f$ denotes the T-even final state interactions and $D_t$ is the T-odd contribution, which gives the fundamental violation of the time-reversal symmetry. At the lowest order in the electroweak corrections $D=0$ and the SM corrections due to the Kobayashi-Maskawa phase and the $\bar{\theta}$ term have been computed in Ref.~\cite{Herczeg:1997se}. Using the current experimental limit for $\bar{\theta}$, the contributions to the $D$ coefficient due the CKM phase and $\bar{\theta}$ are of the order $\sim10^{-12}$ and $\sim10^{-15}$, respectively. Instead, the final state interactions have been computed in Ref.~\cite{Ando:2009jk} and are at the level of $D_f\sim 10^{-5}$ with a $1\%$ accuracy. Keeping in mind the current experimental limit coming from the neutron decay~\cite{Mumm:2011nd,Mumm:2011zz}
\begin{equation}
D_n =( -1.0\pm2.1)\times 10^{-4}, 
\end{equation}
there still a window for new physics observation up to $D_n \sim 10^{-7}$. 

This constraint was previously studied in the context of the mLRSM in Ref.~\cite{Ng:2011ui}, where it was argued that EDM systems already exclude the possibility of probing CP violation within the mLRSM model in $\beta$-decays since $D_t\lesssim10^{-7}$ from the neutron EDM. In this paper, we show that this conclusion changes in the light of two facts. The first is that the chiral effective field theory sensitivity of $d_n$ to the mLRSM parameter is smaller than the one reported in Ref.~\cite{Ng:2011ui}, as concluded in Refs.~\cite{Engel:2013lsa,Seng:2014pba}. The other is that our global fit gives an upper bound to the mLRSM parameter. It turns out that in the mLRSM, the same dim-6 operator in Eq.~\eqref{dim6} that generates the EDM also contributes to the $D$ coefficient in $\beta$-decay. One may thus express (for example) $d_n$ in terms of ${\bar\theta}$ and $D_t$ as 
\begin{equation}
\label{eq:dnDt}
d_n = \alpha_n \bar{\theta}+ \beta_n^{\varphi ud} \left(\frac{D_t}{\kappa}\right)\,,
\end{equation}
where 
\begin{equation}
\frac{D_{t}}{\kappa} = \kappa_{LR} 
\end{equation}
and $\kappa\simeq 0.87$ for the neutron~\cite{Ng:2011ui}. Using the result shown in Fig.~\ref{CHI2} it is easy to see from our global fit that $\left| \frac{D_{t}}{\kappa}\right|\leq 2.0\times 10^{-5}$ at $95\%$ C.L, which is well above the uncertainty associated with $D_f$. Thus, a future, more sensitive probe of the T-odd, triple correlation in $\beta$-decay would probe a portion of the mLRSM parameter space not currently constrained by EDM searches.

\section{Conclusions}\label{conclusions}
In this paper, we performed a global fit to limit the $\bar{\theta}$ parameter and the new physics contribution in the context of the minimal left-right symmetric model. We found that the ``sole-source" limit of $\bar{\theta}\simeq 10^{-10}$ gets relaxed by roughly one order of magnitude from our global analysis. We then discuss the implications for both parity and charge conjugation as the left-right symmetry. This new limit cannot be evaded by invoking cancellations between the $\bar{\theta}$ contribution and the new physics one. Finally, we revisit the connection with the $D$ coefficient of $\beta$-decay in the light of our global fit and find that there is still room for the observation of fundamental time-reversal symmetry violation in $\beta$-decay experiments. 

\section*{Appendix}

Here, we show two figures giving the correlation between $\chi^2$ and the nuclear parameter $a_{1}$ for mercury and radium.
\begin{figure}[h]
 \centering
 \includegraphics[width=.9\columnwidth]{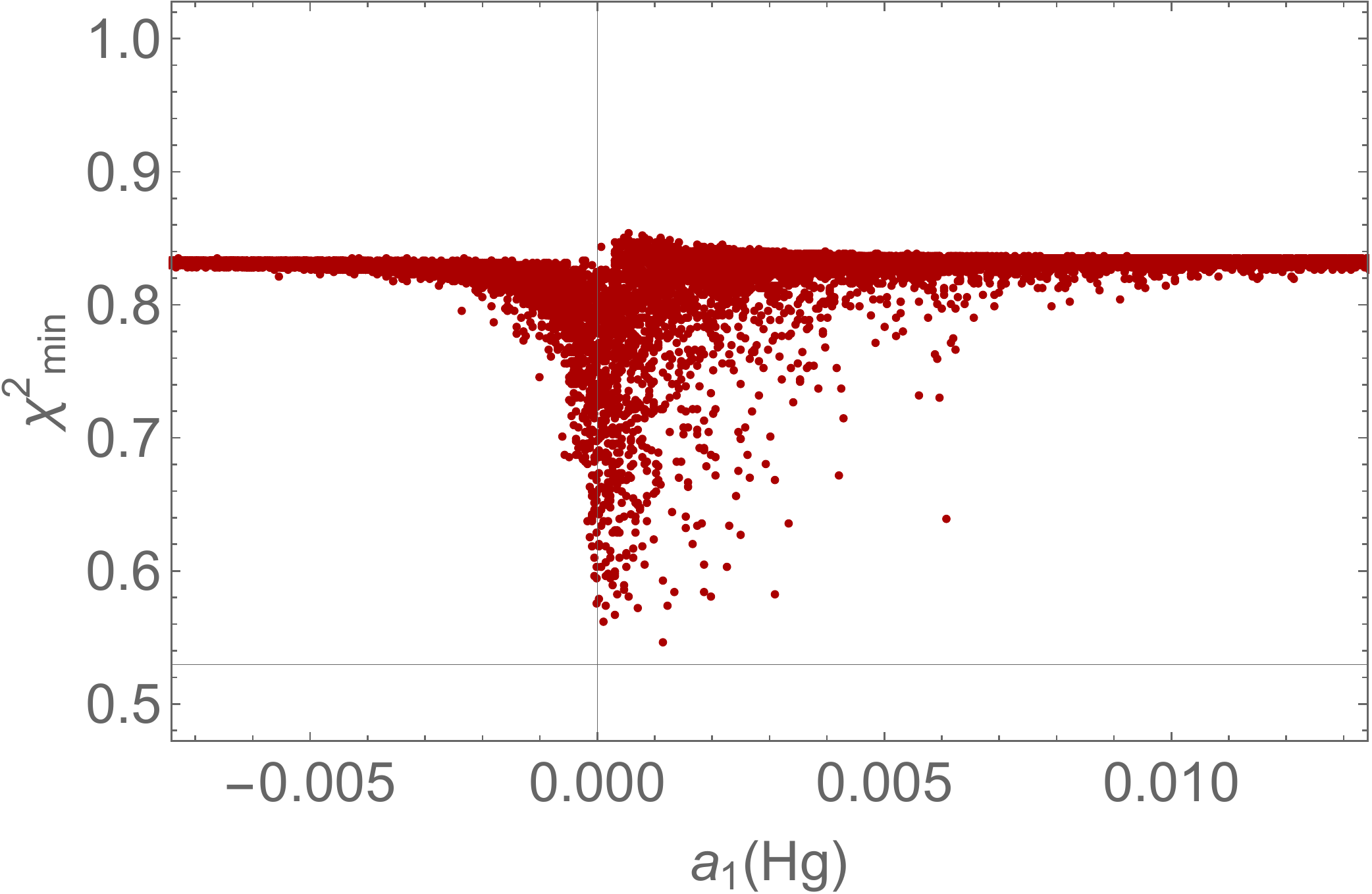}
 \caption{Values of $\chi^2_{min}$ of the $a_1(\mathrm{Hg})$ nuclear parameter. }
 \label{CHI2VSa1Hg}
\end{figure}
\begin{figure}[h]
 \centering
 \includegraphics[width=.9\columnwidth]{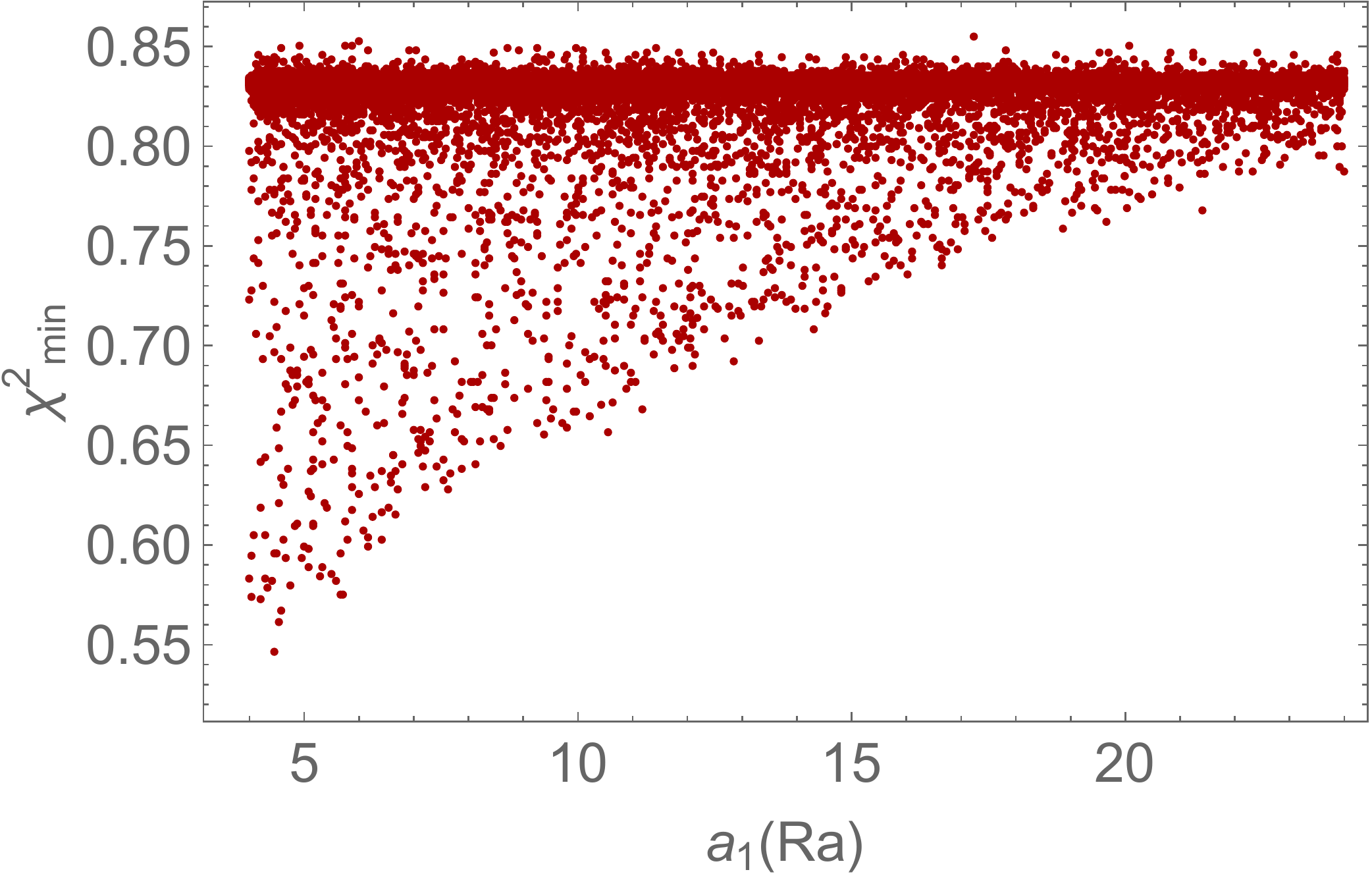}
 \caption{Values of $\chi^2_{min}$ of the $a_1(\mathrm{Ra})$ nuclear parameter. }
 \label{CHI2VSa1Ra}
\end{figure}

\section*{Acknowlegdements}

JCV thanks Goran Senanovi\' c for encouragement and useful and illuminating discussions. JVC would like to thank T. Chupp for correspondence about the global fit and Jordy De Vries for many valuable discussions and a careful reading of the manuscript.  JCV was supported in part under the U.S. Department of Energy contract DE-SC0015376. MJRM was supported in part under the U.S. Department of Energy contract DE-SC0011095. MJRM was also supported in part under National Science Foundation of China grant No. 19Z103010239.

\section*{References}

\bibliography{EDMLR}

\end{document}